\documentclass[12pt]{iopart}
\usepackage{iopams}  
\usepackage{epsfig}
\usepackage{verbatim}
\usepackage{hyperref}
\usepackage[square,sort&compress,numbers]{natbib}
\usepackage{url}
\begin{document}

\title[Tuning into Scorpius X-1]{Tuning into Scorpius X-1: adapting a continuous gravitational-wave search for a known binary system}

\author{Grant David Meadors$^1$, Evan Goetz$^2$, Keith Riles$^2$}

\address{$^1$Max-Planck-Institut f\"{u}r Gravitationsphysik, Am M\"{u}hlenberg 1, 14476 Potsdam and Callinstra{\ss}e 38, 30167 Hannover, Germany}
\address{$^2$University of Michigan, 450 Church Street, Ann Arbor, Michigan 48109, USA}
\ead{grant.meadors@aei.mpg.de}
\vspace{10pt}
\date{\today}

\begin{abstract}
We describe how the TwoSpect data analysis method for continuous gravitational waves (GWs) has been tuned for directed sources such as the Low Mass X-ray Binary (LMXB), Scorpius X-1 (Sco X-1).
A comparison of five search algorithms generated simulations of the orbital and GW parameters of Sco X-1. 
Where that comparison focused on relative performance, here the simulations help quantify the sensitivity enhancement and parameter estimation abilities of this directed method, derived from an all-sky search for unknown sources, using doubly Fourier-transformed data.
Sensitivity is shown to be enhanced when the source sky location and period are known, because we can run a fully-templated search, bypassing the all-sky hierarchical stage using an incoherent harmonic sum.
The GW strain and frequency as well as the projected semi-major axis of the binary system are recovered and uncertainty estimated, for simulated signals that are detected.
Upper limits on GW strain are set for undetected signals.
Applications to future GW observatory data are discussed.
Robust against spin-wandering and computationally tractable despite unknown frequency, this directed search is an important new tool for finding gravitational signals from LMXBs. 
\end{abstract}

%
%
%
%
%

\section{Introduction}
\label{binary_NS}

Continuous gravitational waves (GWs) from neutron stars in binary systems seem likely to be one of the most interesting types detectable by ground-based interferometric observatories.
Binary systems constitute 237 of 578 (44\%) of known pulsars in the ATNF catalog (v1.53, 2015)~\cite{ManchesterATNF2005} with rotational frequency faster than 5 Hz, the approximate lower bound of the frequency range of Advanced LIGO and Advanced Virgo~\cite{HarryALIGO2010,Acernese2009}.
TwoSpect~\cite{GoetzThesis,GoetzTwoSpectMethods2011} is one data analysis method for targeting systems of Low Mass X-ray Binaries (LMXBs), including the prototypical Scorpius X-1 (Sco X-1).
This method can apply to continuous-wave (CW) GWs of unknown frequencies from neutron stars in binary systems, with unknown sky locations, orbital periods, or projected semi-major axes.
It was used in a prior all-sky search for CWs from unknown sources with data from LIGO Science Run 6 and Virgo Science Runs 2 \& 3~\cite{GoetzTwoSpectResults2014}.
This paper describes analysis modifications, when sky location and orbital period are known, to run a \textit{directed} search.
The relative performance of this method was shown in a comparison of five algorithms~\cite{ScoX1MDC2015PRD}.
More sensitive than the all-sky search, the new directed methodology is presented here, including discussion of detection criteria, parameter estimation, and upper limit strategy.
Now, in the wake of the first GW observation~\cite{GW150914LIGO}, we soon hope to observe LMXBs such as Sco X-1.

CW analyses for isolated neutron stars are computationally-demanding~\cite{Brady1998} but conceptually simple.
Ellipticity in a rotating star, or stellar $r$-modes~\cite{Shawhan2010,Owen2010}, would generate a time-varying mass quadrupole moment that could induce CW emission. 
When these emissions, with a dimensionless strain amplitude $h_0$ and a phase evolution described in Section~\ref{signal-model-section}, arrive at a GW observatory, they may be very faint. 
CW analyses include the $\mathcal{F}$-statistic, Hough, StackSlide and PowerFlux methods~\cite{Jaranowski1998,HoughTransformKrishnan2004,LSCPulsarS4,LSCPowerFlux2009,PowerFluxMethod2010,PowerFluxAllSky2012}.
Many stars, possibly the best GW sources, are unknown or have ephemerides insufficiently precise to make a fully-coherent search tractable.
Different strategies are therefore used depending on available information.

CW searches are categorized as \textit{all-sky} (for unknown objects), \textit{directed} (sky location known) and \textit{targeted} (spin frequency also known).
When sky location and other ephemerides are known,
computational cost can often be reduced or reinvested in increased sensitivity.
As recently reviewed~\cite{Riles2013},
directed~\cite{WetteCasA2008,AbadieCasA2010} and targeted~\cite{DupuisWoan2005,AasiPulsarInitialResults2014} searches now exist.

Signals from neutron stars with rotational periods of milliseconds should populate the GW spectrum.
Millisecond pulsars appear to have a speed limit somewhat higher than 700 Hz but below their expected relativistic break-up speed~\cite{Chakrabarty2003}: GW emission is a possible cause.
Dedicated analyses are motivated by the large fraction of these pulsars that are in binary systems. 

Binary systems intrinsically have more parameters to analyze than isolated stars. 
TwoSpect~\cite{GoetzTwoSpectMethods2011} uses doubly Fourier-transformed data in the first practical all-sky search for unknown neutron stars in binary systems~\cite{GoetzTwoSpectResults2014}.
Other binary searches include developments of the Sideband~\cite{Messenger2007CQG,Sammut2014PRD}, Radiometer~\cite{Ballmer2006CQG,AbadieStoch2011}, Polynomial~\cite{2010JPhCS.228a2005V}, and CrossCorr~\cite{Dhurandhar2008,ScoX1CrossCorr2015PRD}. 
A systematic comparison~\cite{ScoX1MDC2015PRD} challenged 
these five methods
to detect simulated signals from Sco X-1.
A stacked $\mathcal{F}$-statistic, derived from the coherent $\mathcal{F}$-statistic~\cite{AbbottScoX12007}, is under investigation~\cite{LeaciPrixDirectedFStatPRD}.

LMXBs, including Sco X-1, are believed to \textit{spin-up} (increase rotational frequency) by accretion-driven recycling~\cite{PapaloizouPringle1978}.
Accretion can also lead to non-axisymmetry that induces emission.
These mechanisms suggest a prime GW source.
In the \textit{torque balance} hypothesis, spin-up would continue until it equaled and canceled \textit{spin-down} from GW emission~\cite{Wagoner1984}.
This hypothesis can be quantified in terms of GW strain.

Torque balance predicts a characteristic strain $h_c$ given by Equation~\ref{torque_bal_eq} (Equation 4 of Bildsten~\cite{Bildsten1998}; note $h_c/h_0 = 2.9/4.0$).
For an LMXB with flux $\mathcal{F}_\mathrm{X-ray}$ and NS that rotates at a frequency $\nu_s$, radiating at the quadrupolar GW frequency $f = 2\nu_s$,

\begin{equation}
h_c \approx 4\times10^{-27}\left(\frac{300\mathrm{~Hz}}{\nu_s}\right)^{1/2}\left(\frac{\mathcal{F}_\mathrm{X-ray}}{10^{-8}\mathrm{~erg~cm}^{-2}\mathrm{~s}^{-1}}\right)^{1/2}.
\label{torque_bal_eq}
\end{equation}

With X-ray flux ($3.9 \times 10^{-7}$ erg cm$^{-2}$ s$^{-1}$~\cite{Watts2008}) but unknown frequency for Sco X-1, this limit can be evaluated for $f$ between 50 and 1500 Hz~\cite{ScoX1MDC2015PRD}, although large systematic uncertainties mean $h_0$ could be larger or smaller:

\begin{eqnarray}
h_0 &= \frac{4.0}{2.9} h_c, \nonumber \\
 &\approx 3.5\times10^{-26}\left(\frac{600\mathrm{~Hz}}{f}\right)^{1/2}, \nonumber\\
&\rightarrow
  h_0 < 2.2\times10^{-26}~(f=1500~\mathrm{Hz}),~h_0 <1.2\times10^{-25}~(f=50~\mathrm{Hz}).
\label{ScoX1_torque_bal}
\end{eqnarray}

Angular momentum from accretion would counterbalance that lost to GW emission. 
If true, $f$ would remain stable for long durations, aside from spin-wandering due to variations in accretion.
Computational costs are thus reduced
by the restricted range of the frequency-derivative.

The torque-balance levels predicted by Equation~\ref{ScoX1_torque_bal} are near the estimated thresholds of detection, meriting effort in enhancing the sensitivity of our methods.
Directed search improvements skip the computation-saving hierarchical steps of the all-sky search, testing all points in parameter space with templates for enhanced sensitivity to a signal from a known sky location.
This paper details the directed, fully-templated method and shows its application to simulated data containing Sco X-1 signals.

\section{Signal model}
\label{signal-model-section}

GW signals are defined by $h(t)$, strain as a function of time.
Models of $h(t)$ in a binary search~\cite{GoetzTwoSpectMethods2011} depend on phase evolution, $\Phi(t)$,

\begin{equation}
\Phi (t) = \Phi_0 + 2 \pi f_0 \cdot \tau(t) + \Delta f_\mathrm{obs} \cdot P \cdot \sin \left(\Omega [t - T_\mathrm{asc}] \right),
\end{equation} 
\begin{equation}
h(t) 
= h_0 F_+ \frac{1+\cos^2 \iota}{2}\cos \Phi(t) +
  h_0 F_\times \cos \iota \sin \Phi(t).
\end{equation}

\noindent 
Here solar-system barycentered time is $\tau(t)$ and $\Phi(t=0) \equiv \Phi_0$. 
We also neglect spin-wandering, which could manifest as stochastic variation in $\Phi_0$.

The TwoSpect model does not currently search over \textit{amplitude parameters}: $h_0$ is the GW strain amplitude, $\iota$ is the inclination angle of the neutron star with respect to the source, $\psi$ is the GW polarization angle.
While $h_0$ can be recovered, its estimation is confounded by $\iota$.
Both $\iota$ and $\psi$ affect amplitude through detector response, which depends on angle via the plus- and cross antenna functions, $F_+$ and $F_\times$.
Initial GW phase, $\Phi_0$, further specifies the signal, but is neither explored nor recoverable.

Our search is over unknown \textit{Doppler parameters}, which drive signal evolution.
Sky location $(\alpha, \delta)$ is known for Sco X-1, as is
orbital period $P = 2\pi / \Omega$.
Initial orbital phase is fixed by time of ascension $T_\mathrm{asc}$, to which our method is currently insensitive.
Frequency $f$ and projected semi-major axis $a \sin i$ (expressed in light-seconds, ls)  must be searched over.
Because the signal is frequency-modulated by the orbital motion of the source in the circular binary system, the latter parameter manifests through modulation depth $\Delta f_\mathrm{obs} = 2 \pi f_0 \cdot (a \sin i) / (c P)$.

\section{Analysis statistic}
\label{twospect-statistics}

TwoSpect has already been described for the all-sky analysis~\cite{GoetzTwoSpectMethods2011,GoetzTwoSpectResults2014}. Only a brief summary is given here on the elements common to the directed analysis, with details in~\ref{mathematical-appendix-methods}.

Data containing calibrated GW strain, originally recorded by each observatory as time series, are read into the program from a sequence of short Fourier transforms (SFTs).
These SFTs, on the order of minutes long, are shorter than the total observation time $T_\mathrm{obs}$, on the order of months.
Noise and antenna pattern weights are applied to enhance sensitivity by weighting those SFTs that are more sensitive to a putative source. 
Each SFT contains $K$ frequency bins.
The Earth's motion Doppler shifts the apparent frequency of a source, so these bins must be barycentered: their indices are shifted such that an unmodulated frequency from a given sky location remains in the same bin for all SFTs.
These frequency bins represent the instantaneous frequency $f$ of the signal.
A second Fourier transform is performed, transforming the power in each frequency bin: the transform is from SFT time to $f'$.
Effectively, $f'$ represents the orbital frequency of the binary system. 

Templates are the $M$ pixel weights $w_i$ in the 2-D image-plane $(f, f')$.
Each template corresponds to a signal model from the $(f_0,\Delta f_\mathrm{obs})$ astrophysical parameter space.
After the second Fourier transform, the powers $Z_i$ of data pixels in the $(f, f')$ are measured and noise-background $\lambda_i$ estimated.
The test statistic, $R$, is the projection of the data (noise-subtracted powers) onto the templates, normalized by the templates:

\begin{equation}
R=\frac{\sum_{i=0}^{M-1}w_{i}[Z_{i}-\lambda_{i}]}{\sum_{i=0}^{M-1}[w_{i}]^{2}}.
\label{TwoSpect_R_statistic}
\end{equation}

Estimated $p$-values for this statistic allow determination of detection probability.
Expositions of the detection statistic and signal model can be found in previous methods papers~\cite{GoetzTwoSpectMethods2011,TwoSpectCoherentGoetz2015arxiv}.

\section{Application to directed searches}
\label{application_to_directed_searches}

\subsection{Sensitivity and computational cost}

The original design of the all-sky analysis employs hierarchical analysis to control computational costs while still maintaining good sensitivity to a broad parameter space~\cite{GoetzTwoSpectMethods2011}.
Each narrow frequency band requires corrections for the Doppler effect caused by Earth's motion. 
Because this correction depends on sky position, an all-sky search for typical SFT lengths of several minutes would need $\mathcal{O}(10^{18})$ templates; in practice, an incoherent harmonic sum is used to reduce this number, only calculating templates for candidates with significant sums.
This practice limits sensitivity~\cite{GoetzTwoSpectMethods2011}.
The new, directed search in this paper can use $\mathcal{O}(10^{8})$ templates at a single sky location with full sensitivity, because 
$R$-statistics are returned for every template.

For Sco X-1, distance, eccentricity, X-ray luminosity, sky location, and orbital period are known with good precision and projected semi-major axis is measured to be $1.44\pm0.18$ ls, (Table~\ref{scox1_table_params}). 
NS spin frequency, however, is unknown~\cite{Galloway2014}.
Other targets, such as XTE J1751-305~\cite{Markwardt2002}, have known frequency, reducing computational costs substantially.
Sco X-1, XTE J1751-305, and other LMXBs are the principal sources for the new, directed method, described next.

\Table{
\label{scox1_table_params}
Sco X-1 prior measured parameters from electromagnetic observations.
Note that the projected semi-major axis value is derived from a velocity amplitude of 
$K_1=40\pm5\ {\rm km\,s^{-1}}$~\cite{AbbottScoX12007,ScoX1MDC2015PRD}.
}
\br
Sco X-1 parameter & Value & Units\\
\mr
Distance~\cite{Bradshaw1999} & $2.8\pm0.3$ & kpc\\
Eccentricity ($\epsilon$)~\cite{ScoX1MDC2015PRD} & $< 0.068$ $(3 \sigma)$ & ---\\
Right ascension ($\alpha$)~\cite{2mass06} & 16:19:55.067 $\pm 0.06'' $ & --- \\
Declination ($\delta$)~\cite{2mass06} & $-15^\circ 38'25.02''\pm 0.06''$ & ---\\
X-ray flux at Earth ($\mathcal{F}_\mathrm{X-ray}$)~\cite{Watts2008} & $3.9\times10^{-7}$ &  erg cm$^{-2}$ s$^{-1}$\\
Orbital period ($P$)~\cite{Galloway2014} & $68023.70 \pm 0.04$ & s\\
Projected semi-major axis ($a \sin i$)~\cite{2002ApJ...568..273S} & $1.44\pm0.18$ & ls\\
\br
\endTable

Searching over projected semi-major axis in the range $[-3\sigma_{a \sin i}, +3\sigma_{a \sin i}]$, around the measured value, and over GW signal frequency $f$ from $f_\mathrm{min}$ to $f_\mathrm{max}$, incurs a predictable computational cost.
With a fixed rectangular parameter spacing of $1/(2 T_\mathrm{coh})$ in $f$ and $1/(4 T_\mathrm{coh})$ in $\Delta f_\mathrm{obs}$, using SFTs of coherence time $T_\mathrm{coh}$ and analysis bands of width $f_\mathrm{bw}$, the search uses $N_\mathrm{template}$ templates per observatory (\ref{mathematical-appendix-methods}):

\numparts
\begin{eqnarray}
\fl N_{\mathrm{{template}}} &= \left[1+2f_\mathrm{bw}T_\mathrm{coh}\right] \times
 {\displaystyle \Sigma}_{j=1}^{j = \frac{f_\mathrm{max} - f_\mathrm{min}}{f_\mathrm{bw}}} \left[ 1 + 2\pi\left(f_\mathrm{min} + j f_\mathrm{bw}\right) \frac{4 T_\mathrm{coh}}{cP} 6 \sigma_{a \sin i} \right], \\
\fl &= 2 \left(T_\mathrm{coh} + \frac{1}{f_\mathrm{bw}}\right) \left[ 1+\frac{4 \pi T_\mathrm{coh}}{cP} (6\sigma_{a \sin i})(f_\mathrm{max} + f_\mathrm{min} + f_\mathrm{bw})\right] (f_\mathrm{max} - f_\mathrm{min}).
\label{N_template_simple}
\end{eqnarray}
\endnumparts

\subsection{Participation in a Sco X-1 mock data challenge}

Several analyses can search for CWs from known neutron stars in binary systems.
TwoSpect can also seek unknown systems\footnote{Other all-sky binary search programs are based on the Polynomial~\cite{2010JPhCS.228a2005V} and Radiometer~\cite{AbadieStoch2011} methods.}.
The Sco X-1 Mock Data Challenge (MDC) compares five methods~\cite{ScoX1MDC2015PRD}, including the new, directed version of TwoSpect.
While the MDC published the results of the participants, this paper explains how our results were obtained and how similar analyses apply to forthcoming observations.  

The MDC simulates 50 \textit{open}, unblinded and 50 \textit{closed}, blinded signals,  with signal parameters drawn from Sco X-1 astrophysical priors.
Each signal is also known as an \textit{injection}.
Three observatories are simulated: H1 (Hanford), L1 (Livingston), and V1 (Virgo).
Within $100$ frequency bands of 5 Hz each, $f$ is uniformly distributed.
Bands cover a range from 50~to 1500~Hz and contain Gaussian noise simulating a detector with noise floor $4\times 10^{-24}$~Hz$^{-1/2}$, the expected Advanced LIGO minimum.
Projected semi-major axis $a \sin i$ is Gaussian-distributed with mean $1.44$~ls and standard deviation $0.18$~ls.
Period $P$ is Gaussian-distributed with mean $68023.70$~s and standard deviation $0.04$~s.
Sky location is $(\alpha,\delta) = ($16:19:55.067$,-15^\circ38'25.02'')$.
The uncertainty in period
is small enough that it is not a free parameter.

Our directed search over $f$ and $a \sin i$, with a fixed period, requires approximately $N_\mathrm{template} = 5.0\times 10^7$ templates per observatory, covering 100 bands of 5 Hz each, 500 Hz total. 
With three observatories, $1.5\times 10^8$ templates are needed. 

\begin{figure}
\begin{center}
\includegraphics[width=0.55\paperwidth,keepaspectratio]{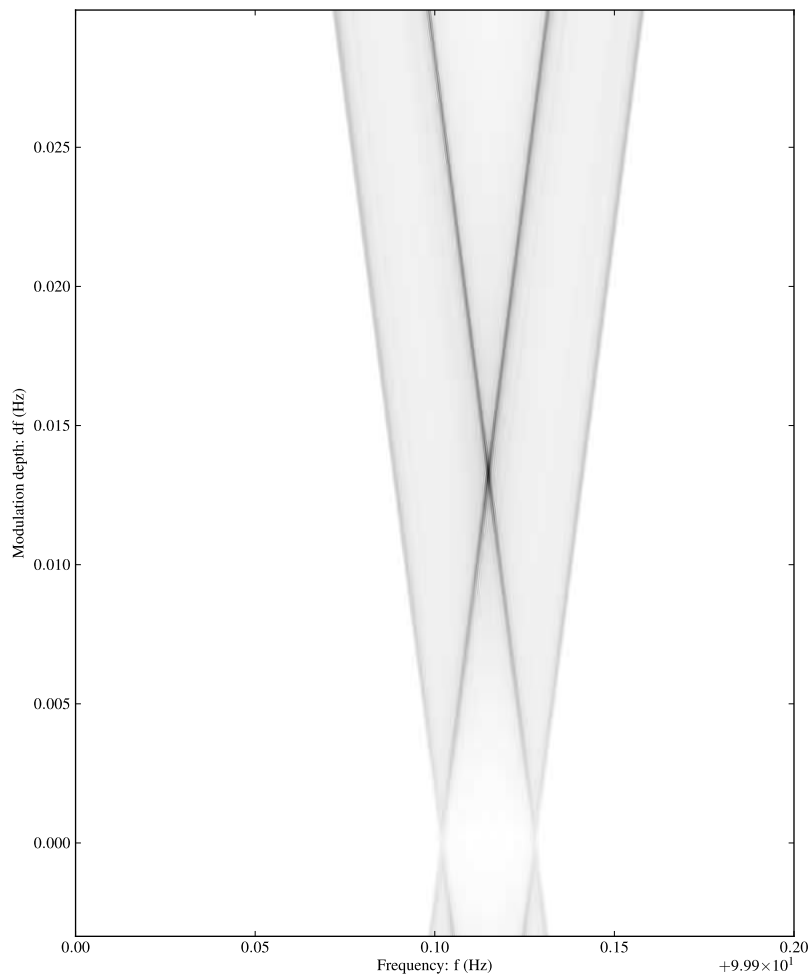}
\caption{
Simulated signal, 100.015 Hz frequency ($f$), 1.44 ls projected semi-major axis (modulation depth 0.0133 Hz, $\Delta f_\mathrm{obs}$), showing single-template $-\log_{10} p$-value.
The graph spans $241$ templates in $\Delta f_\mathrm{obs}$, $721$ in $f$, twice as dense in $\Delta f_\mathrm{obs}$ as in $f$.
A signal with $h_0 = 4\times 10^{-21}$ is injected into a Gaussian noise amplitude spectral density of $4 \times 10^{-24}$ Hz$^{-1/2}$, observed for $10^6$ seconds.  
The $p$-value is extrapolated from the $R$ statistic.
For illustration, $\Delta f_\mathrm{obs}$ extends below zero to show that the code is well-behaved and that the algorithm gracefully mirrors results for negative input parameters.
Most importantly, $-\log_{10}p$ is maximal at the true parameters.
Signal is partly recovered when template $f$ differs from the true signal by plus or minus $\Delta f_\mathrm{obs}$, and also when template $f$ and $\Delta f_\mathrm{obs}$ differ equally from the true parameters.
}
\label{scox1-wide-heatmap-008}
\end{center}
\end{figure}

\begin{figure}
\begin{center}
\includegraphics[width=0.50\paperwidth,keepaspectratio]{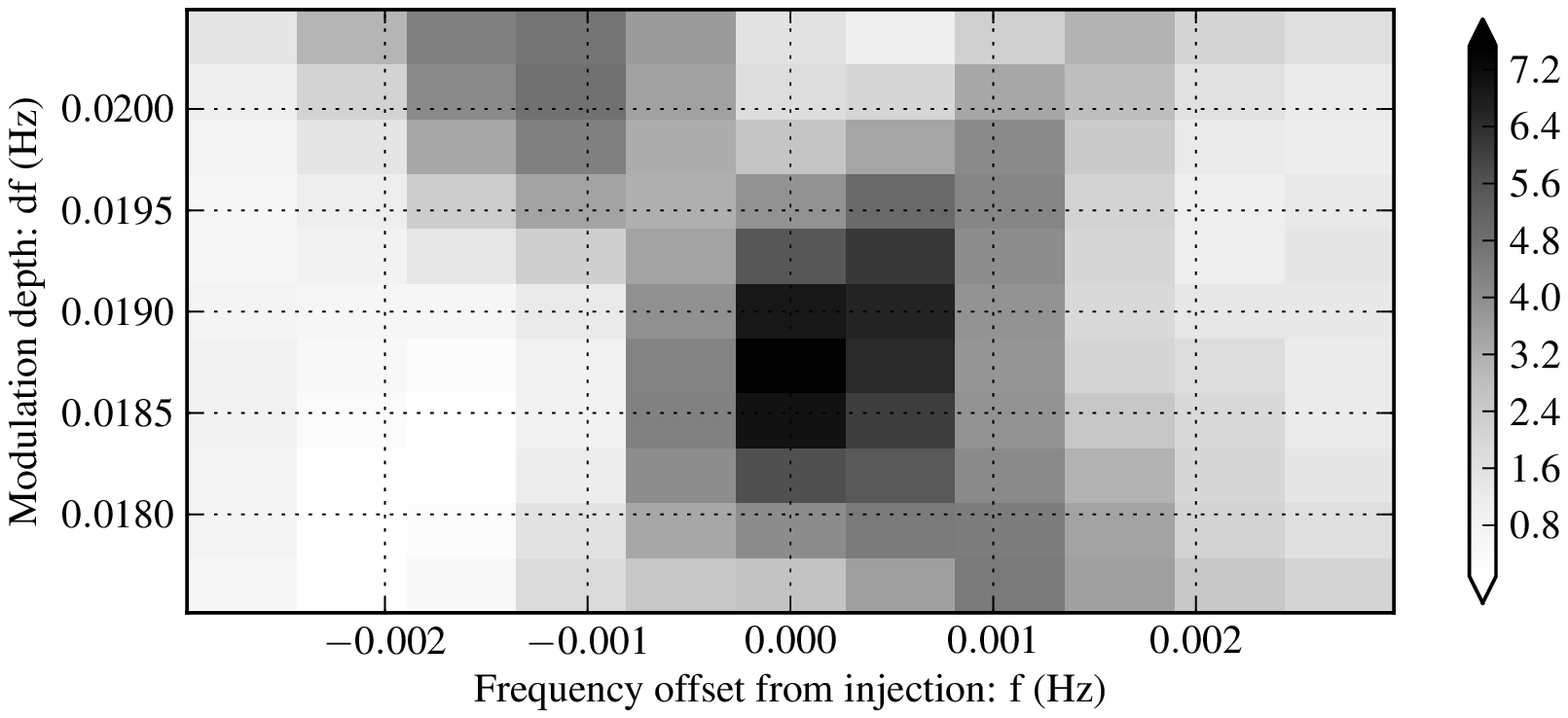}
\includegraphics[width=0.50\paperwidth,keepaspectratio]{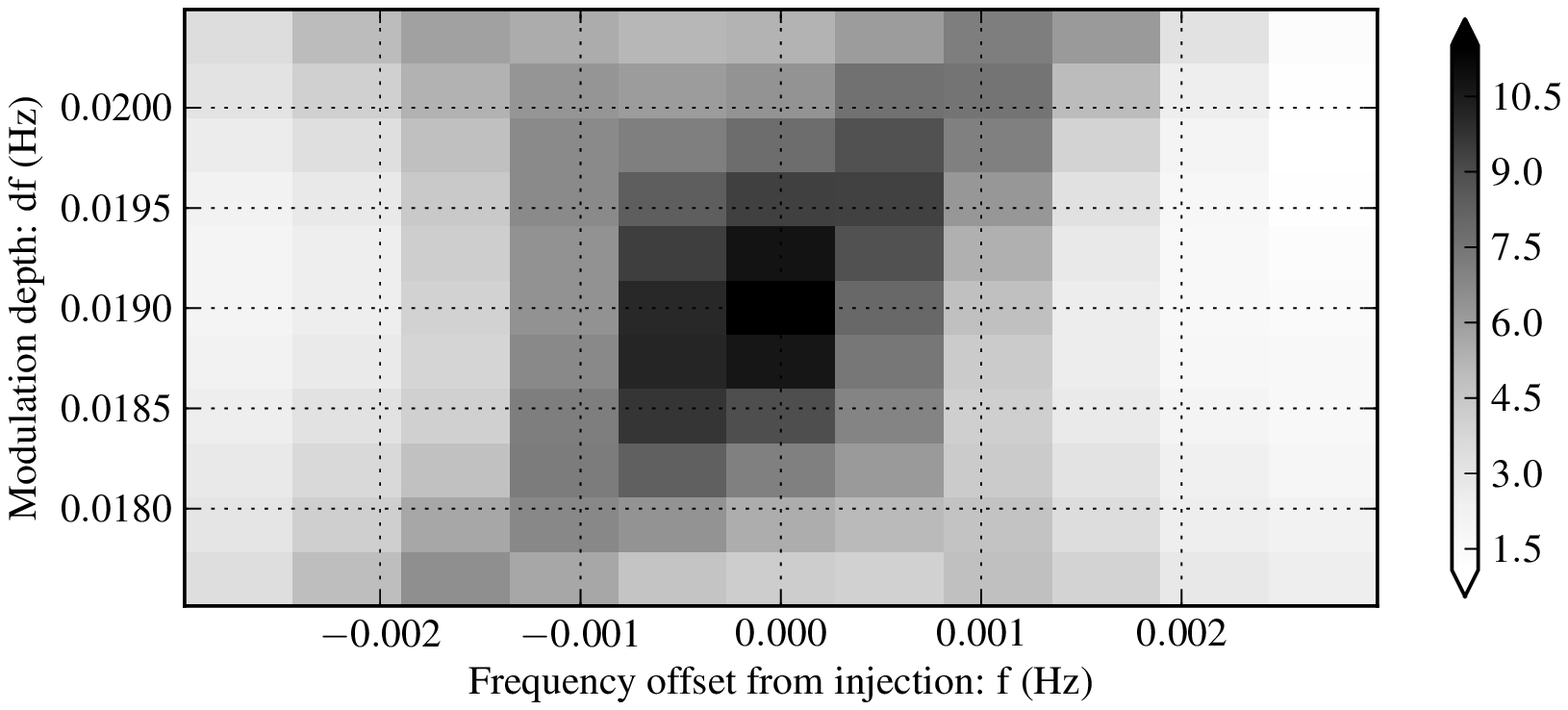}
\includegraphics[width=0.50\paperwidth,keepaspectratio]{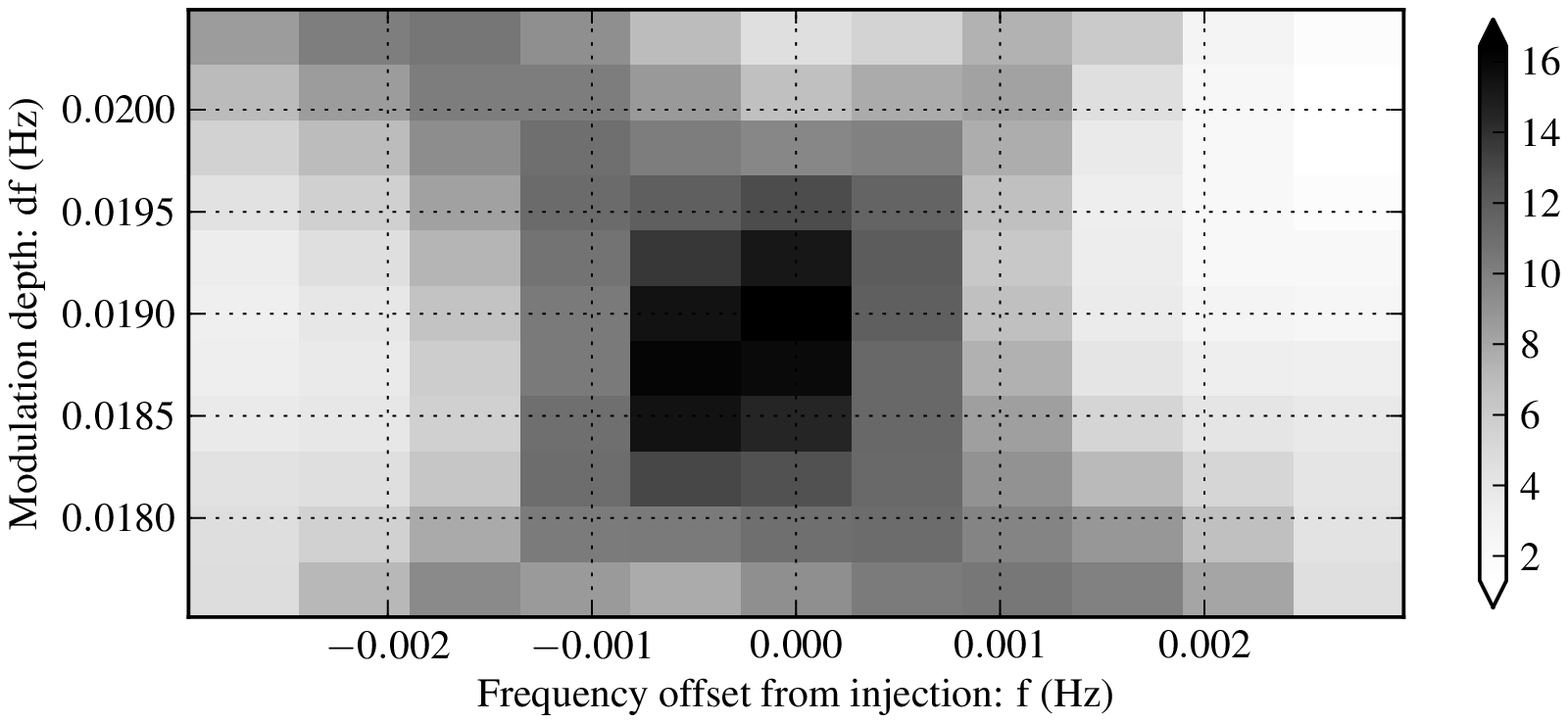}
\caption{Heatmaps for H1, L1, and V1 observatory (top to bottom), of $-\log_{10} p$-value for 
11x11 templates centered on
Sco X-1 MDC signal 8.
This injection was detectable in a year of simulated data at $h_0 = 5.6 \times 10^{-25}$ in noise of $4 \times 10^{-24}$ Hz$^{-1/2}$ and $\cos \iota = 0.09$.
Maxima for all observatories are within a template of the true parameters. 
}
\label{scox1-narrow-heatmap-008}
\end{center}
\end{figure}

Figure~\ref{scox1-wide-heatmap-008} shows a wide analysis that is magnified in Figure~\ref{scox1-narrow-heatmap-008}.
These analyses require minimal modification from that of the all-sky search: $R$-statistic and $p$-value code was identical, aside from looping over all template positions in the parameter grid. 

\section{Directed search demonstration and outlier follow-up}
\label{demonstration_of_the_method}

The new, directed method is demonstrated through the MDC described in Section~\ref{application_to_directed_searches}.
This MDC paper~\cite{ScoX1MDC2015PRD} presents the performance of each participating pipeline but not the details of our method.
Here we highlight methodology not elaborated in the MDC paper, referencing the performance of the directed method for illustration and for contrast with the all-sky method~\cite{GoetzTwoSpectMethods2011}.

After writing a loop over templates, post-processing is the main modification to the pipeline: new detection criteria and techniques for parameter and upper limit estimation are required, because the fully-templated output differs significantly from that of hierarchical method.
In this section, we detail how these results have been obtained and establish a basis for future applications.

\subsection{Overview of detection and parameter estimation}

A set of extremal $p$-value outliers in 5 Hz bands is produced for each 
observatory, subject to a $p$-value threshold inferred from Gaussian noise.
These sets are compared in pairwise coincidence (H1-L1, H1-V1, or L1-V1).
Coincidence of outliers allows $f$ or $\Delta f_\mathrm{obs}$ to differ by $\leq 1/T_\mathrm{coh}$ between observatories, based on prior experience from the all-sky search~\cite{GoetzTwoSpectResults2014}.
This allowed difference is $2$ steps in the $f$ grid or $4$ in the $\Delta f_\mathrm{obs}$ grid.
Surviving outliers are classified as detections. 

For a given detection in one band, the signal parameters are inferred from the values of the template with the highest (single-observatory) $p$-value.
These parameters include $f$, $a \sin i$, and $h_0$.
Again, for open signals, the true parameters were known in advance.
Total uncertainty in $f$ and $a \sin i$, and non-systematic uncertainty (random) in $h_0$, is determined
from the standard deviation of the set of parameters of recovered open signals compared to their true parameters, as detailed in section~\ref{param_est}.

The estimated strain is $h_\mathrm{rec} \propto R^{1/4}$.
Systematic uncertainty arises from the unknown inclination angle, $\cos \iota$, which dominates the total uncertainty in strain. 
This ambiguity cannot be resolved with the present algorithm and depends
partially on the assumed prior distribution of signal amplitudes; the
uncertainty is estimated by simulation in Section~\ref{cosiAmbiguity}.
With new algorithm enhancements since the MDC, $\cos \iota$ can become a searched parameter.

Due to the uniform noise floor and low number of injections, 
a single upper limit value is declared for all 100 bands based on the best estimate of the 95\%
 confidence level of non-detected signals in the open set of injections. 

In future applications to real data, the directed analysis can be post-processed using the detection criteria and parameter estimation methods described here.
Improved upper limits methods are underway, and a strong candidate signal would likely be followed-up with other analyses, but the core pipeline is the same.

\subsection{Detection claims}
\label{Detection_criteria_subsection}

Studying the Gaussian noise in the Sco X-1 MDC open data set,
we can set thresholds for detection.
In real searches, detection candidates are followed-up, so the same thresholds developed here are used to mark interesting candidates.

To obtain a Gaussian noise sample, we excise injection signals, which are visible in the $f$ vs $\Delta f_\mathrm{obs}$ plane. 
The excised region depends on injection frequency $f_\mathrm{inj}$ and modulation depth $\Delta f_\mathrm{inj}$, on the Earth-orbital Doppler shift ($\approx 10^{-4}$), and additional bins (at least 10) to avoid spectral leakage.
The half-width excised for each injection is $\delta f_\mathrm{inj}$:

\begin{equation}
\delta f_\mathrm{inj} = 2 \times (10^{-4}\times f_\mathrm{inj} + \Delta f_\mathrm{inj}) + \frac{10}{360}\mathrm{ Hz}.
\end{equation}

\noindent 
The remaining noise sample is the data set with all intervals $[f_\mathrm{inj}-\delta f_\mathrm{inj},~f_\mathrm{inj}+\delta f_\mathrm{inj}]$ removed.
In the remaining noise sample, the estimated $p$-value distribution is not perfectly uniform, due to gaps in the data.
Nonetheless, the noise sample provides a distribution of template $R$-statistics and $p$-values in the absence of signals.
This procedure provides an empirical measure of the estimated $p$-value that corresponds to an actual false alarm probability of 1\% per 5 Hz frequency band.
Taken together, these let us establish detection criteria.

If there is any candidate surviving the following criteria in a 5 Hz band, we mark it detected, else not detected:

\begin{itemize}
\item single-IFO candidates are the top 200 most extreme $p$-value outliers in a 5-Hz band, of those that pass a $\log_{10}p \leq \mathcal{T}$ threshold, where $\mathcal{T} = -7.75$ if $f <$ 360.0 Hz (those that used 840-s SFTs) or $-12.0$ if $f \geq$ 360.0 Hz (those that used 360-s SFTs).
Note: 
The large discrepancy between the $p$-value thresholds in the MDC is a historical artifact from a configuration error.
The discrepancy is much reduced when this is fixed, as done for future analyses.
Our expectation remains that the threshold should be independent of coherence time.
\item each candidate must survive at least one double-IFO coincidence test, involving a pairwise comparison of single-IFO candidates to see whether they are within 1/$T_{\mathrm{SFT}}$ in both frequency ($f$) and modulation depth ($\Delta f_\mathrm{obs}$).
\end{itemize}

\subsection{Parameter estimation and uncertainty for detected signals}
\label{param_est}

Each template is associated with a particular $(f,\Delta f_\mathrm{obs})$, so parameters are currently read off from the template with the extremal $p$-value corresponding to a detection.
In the future, accuracy might be improved using interpolation, but the MDC validates that the existing method is highly-accurate.
If signals were suspected in real data, this procedure, possibly extended with additional simulations, could generate a parameter space volume for follow-ups to examine.

The open set of signals, of which 31 of 50 were detected, are the foundation for understanding parameter estimation uncertainty.
The reconstructed $h_0$ output is $h_\mathrm{rec} = C R^{1/4}$.
The value of $C$ is determined from the mean value of a large number of simulations for circularly-polarized waves over the whole sky and full range of $f$, $P$, and $\Delta f_\mathrm{obs}$.

Then $h_\mathrm{rec}$ is rescaled twice, first by $\rho_R$ for more accurate measurement at the Sco X-1 period and modulation depth, and second by $\rho_\mathrm{\cos \iota}$ for unknown $\cos \iota$.
Thus the final claimed value of $h_0$ for a signal $j$ is $(h_0)_j$:

\begin{equation}
(h_0)_j = \rho_R \rho_\mathrm{\cos\iota} h_{\mathrm{rec},j}.
\end{equation} 

\noindent The first scale factor, $\rho_R = 1.11$, corrects the average values of $h_\mathrm{rec}$ ($h_0$-reconstructed) in the open set to match the corresponding $h_\mathrm{eff}$ ($h_0$-effective, given circular polarization weightings). That is,  $\bar{h}_\mathrm{eff} = 1.11 \times \bar{h}_\mathrm{rec}$, where $h_\mathrm{eff}$ is defined \textit{a priori} by $h_\mathrm{inj}$ ($h_0$-injected) and $\cos \iota$:

\begin{equation} h_{\mathrm{eff}} = \frac{1}{\sqrt{2}}\sqrt{ \left(\frac{1+\cos^2 \iota}{2}\right)^2 + \left(\cos \iota\right)^2 } \times h_{\mathrm{inj}}.
\label{rescale_cosi_eq}
\end{equation}

In the MDC study~\cite{ScoX1MDC2015PRD}, 4 of the 31 detected, open signals account for the largest frequency estimation error.
This error arose from a misconfiguration that does not affect the other analyses; it was addressed by taking those 4 as one class and the remaining 27 as another, a step that should be unnecessary in future analyses.
Then, the uncertainty due to random error for $h_0$, $f$, and $a \sin i$ is estimated by the standard deviation between the recovered and true parameters:

\begin{eqnarray}
\sigma_f^2 &= \frac{1}{N_\mathrm{open}-1} \sum\limits_{j \in \{ \mathrm{open} \} }^{N_\mathrm{open}} \left( f_{\mathrm{rec},j} - f_{\mathrm{inj},j} \right)^2 ,\\
\sigma_{a \sin i}^2 &= \frac{1}{N_\mathrm{open}-1} \sum\limits_{j \in \{ \mathrm{open} \} }^{N_\mathrm{open}} \left( (a \sin i)_{\mathrm{rec},j} - (a \sin i)_{\mathrm{inj},j} \right)^2 ,\\
\sigma_{h_0,\mathrm{rand}}^2 &= \frac{1}{N_\mathrm{open}-1} \sum\limits_{j \in \{ \mathrm{open} \} }^{N_\mathrm{open}} \left( \rho_R\times h_{\mathrm{rec},j } - h_{\mathrm{eff},j} \right)^2 ,
\end{eqnarray}

\noindent where $N_\mathrm{open}$ is the number of open injections, $\sigma_f$, $\sigma_{a \sin i}$, and $\sigma_{h_0}$ are the uncertainties we state for recovered $f_\mathrm{rec}$, $(a \sin i)_\mathrm{rec}$, and $h_\mathrm{rec}$ given injected $f_\mathrm{inj}$, $(a \sin i)_\mathrm{inj}$, and $h_\mathrm{inj}$.

The error between injected and recovered parameters does not show any other clear correlation with $p$-value or signal frequency, at least in the 31 detected signals.
Except for the most marginally detected signals, where noise fluctuations matter, uncertainty in $f$ and $a \sin i$ is dominated by the template grid spacing.
The $\sigma_f$ and $\sigma_{a \sin i}$ error bars have been used uniformly for claiming uncertainties on the signals in the MDC.

The largest source of uncertainty for $h_0$ comes from correction for systematic underestimation, multiplying a factor of $1.74$ into $C$.
This uncertainty is the ambiguity in $\cos \iota$ discussed in Section~\ref{cosiAmbiguity}.
Parameter estimation uncertainty for $f$ and $a\sin i$ is then just the random error; for $h_0$, it is the quadrature sum of random error and $\cos \iota$ ambiguity.

\subsection{Ambiguity from $\cos \iota$}
\label{cosiAmbiguity}

The largest systematic uncertainty in $h_0$ comes from the unknown $\cos \iota$. 
The method is optimized for $|\cos \iota| = 1$ and computes the $R$ statistic by weighting the SFTs assuming circularly-polarized GWs, which still provides good sensitivity for other polarizations. 
Recall that $h_\mathrm{rec}$ must be scaled by $\rho_R=1.11$ to match $h_\mathrm{eff}$.
When a source is circularly polarized, the analysis estimates $\rho_R h_\mathrm{rec} \approx h_\mathrm{inj}$.
In the case of linear polarization, Equation~\ref{rescale_cosi_eq} indicates that $h_\mathrm{inj}$ will be about $2^{3/2}$ times larger than $h_\mathrm{eff}$ (and so $h_\mathrm{rec}$).
The aim is to find an average conversion factor from $h_\mathrm{rec}$ to $h_\mathrm{inj}$ and a robust estimate of the uncertainty.

Since $h_\mathrm{eff}$ of circularly-polarized signals is greater, those signals are more easily detectable than linearly-polarized signals of equivalent $h_\mathrm{inj}$.
Therefore the signals that are detectable are biased, near threshold, to being more likely circularly polarized.
This ``circularizes" the correction factor, depending on the detection efficiency of the pipeline and on the assumed prior distribution of strain amplitudes. 
Although the effect is minor, estimating its size requires simulation.

\begin{figure}
\begin{center}
\includegraphics[trim=0 0 0 0, clip, keepaspectratio,height=0.40\paperheight]{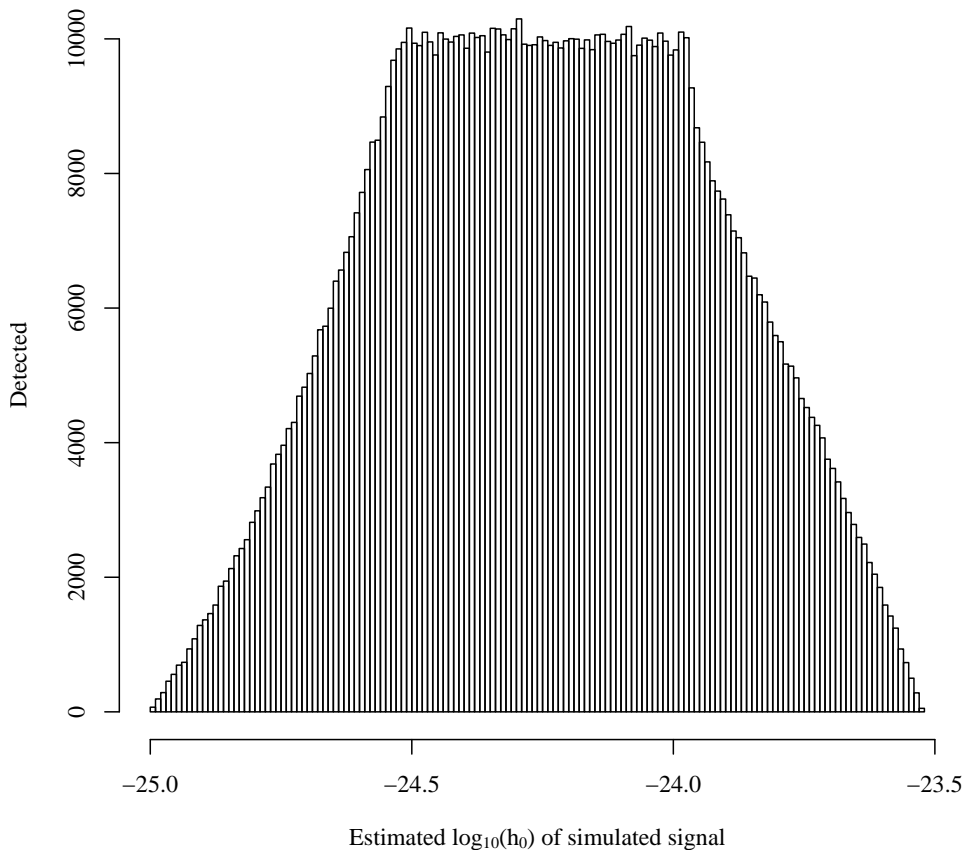}
\caption{Histogram, 150 bins, of the distribution of 2 million simulated signals, strain between $3\times 10^{-26}$ and $3\times 10^{-24}$ under a log-uniform distribution, following application of cos $\iota$ and detection efficiency cuts.
\label{fig:plotheffdisth0detected}}
\end{center}
\end{figure}

The simulation generates 2 million signal amplitudes between $h_\mathrm{inj} = 3\times 10^{-26}$ and $h_\mathrm{inj} = 3\times 10^{-24}$ with a distribution of 1/$h_0$, the assumed prior distribution of $h_0$ values.
This simulation code is independent of TwoSpect and should apply to similar directed searches.
In this simulation, $\rho_R$ is 1 for simplicity, so we can treat $h_\mathrm{rec} = h_\mathrm{eff}$ here.
We model detection efficiency by assuming no signals are detected below $h_\mathrm{eff} = 1\times 10^{-25}$, all are detected above $h_\mathrm{eff}=3\times 10^{-25}$, and the fraction detected is linear in $h_0$ between those values.
Together with a uniform $\cos \iota$ distribution of $[-1, 1]$, this leads to a trapezoidal distribution of recovered, detected $h_0$ values with a curved lower (left) edge (Figure~\ref{fig:plotheffdisth0detected}).

Part of the domain of the simulation must be excluded.
To find the average $h_\mathrm{inj}/h_\mathrm{rec}$ for a given $h_\mathrm{rec}$, every $h_\mathrm{rec}$ must 
correspond to a full sampling of the range of polarizations.
A large $h_\mathrm{inj}$ with linear polarization or small $h_\mathrm{inj}$ with circular polarization could have the same $h_\mathrm{rec}$.
No linearly-polarized signals could produce $h_\mathrm{rec}$ above $1\times 10^{-24}$, because Equation~\ref{rescale_cosi_eq} shows that the largest signal, $h_\mathrm{inj} = 3\times10^{-24}$, would be reconstructed a factor of $2 \sqrt{2}$ smaller, at $h_\mathrm{rec} = 1.06 \times 10^{-24}$ (again, $\rho_R = 1$ for the simulation).
Above $1\times 10^{-24}$, polarizations tend to be more circular, thus the average ratio must exclude this region or it will be biased by the limited range of the simulation $h_\mathrm{inj}$.
Expanding the domain would raise the cutoff, although the resulting ratio would no longer perfectly correspond to the MDC.

With the domain of the simulation determined, we compute that mean ratio of $h_\mathrm{inj}$ to $h_\mathrm{rec}$ to be 1.74. 
Fine-binning $h_\mathrm{rec}$, an interval of $[1.74-\sigma_{\cos \iota},1.74+\sigma_{\cos \iota}]$ encloses 68\% of corresponding $h_\mathrm{inj}$ when $\sigma_{\cos \iota}$ is 0.37 (found by manual optimization).
Therefore our best estimate for the correction factor $\rho_\mathrm{\cos \iota}$, with $\sigma_{\cos \iota}$ inferred as the standard deviation, is $1.74\pm 0.37$. 
This factor multiplies $\rho_R$, which is found to be $1.11$.

The systematic uncertainty, being the uncertainty $\sigma_{\cos\iota}$ in the correction factor, scales with signal strength; the non-systematic (random) is fixed and is also multiplied by the correction factor.
The final estimate of the uncertainty in $h_0$ for signal $j$ is the quadrature sum of the systematic and non-systematic uncertainties:

\begin{equation}
\sigma_{(h_0),j} = \sqrt{\left(\rho_\mathrm{\cos\iota}\times\sigma_{h_0,\mathrm{rand}}\right)^2 + \left(\sigma_\mathrm{\cos\iota}\times \rho_R\times h_{\mathrm{rec},j}\right)^2}.
\end{equation}

For future data, a similar simulation could be run, with an updated detection efficiency model and prior distribution of strains, to find the uncertainty in $h_0$ due to $\cos \iota$ for promising signals.

\subsection{Accuracy of parameter estimation uncertainty claims}

The scheme described above reliably recovers parameters and states uncertainties consistent with the true distribution of errors, as shown in the MDC~\cite{ScoX1MDC2015PRD}.
Verifying the calibration factors and confidence intervals once more, one can confirm that a conservative fraction of $h_0$, $f$ and $a \sin i$ are within their 1-$\sigma$ error bars:
77.4\% for $\sigma_{h_0}$,
74.2\% for $\sigma_{f}$,
and 67.7\% for $\sigma_{a \sin i}$.

\subsection{Upper limits for undetected signals and detection efficiency}

Only upper limits have come from CW searches to date.
Until GWs are detected from neutron stars, the scientific value of a CW analysis is to constrain the plausible $h_0$ and inferred ellipticity from those stars. 
For the MDC-simulated signals, we simplified upper-limit estimation.
In a real detector, noise varies with frequency~\cite{AbadieCalibration2010}; in this simulation, the noise floor is flat at $4\times10^{-24}$ Hz$^{-1/2}$.
Given observational data, we would inject a large number of simulated signals into a number of smaller bands, in order to understand the upper limit as a function of frequency.

\begin{figure}
\begin{center}
\includegraphics[trim=0 0 0 0, clip, height=0.50\paperheight,keepaspectratio]{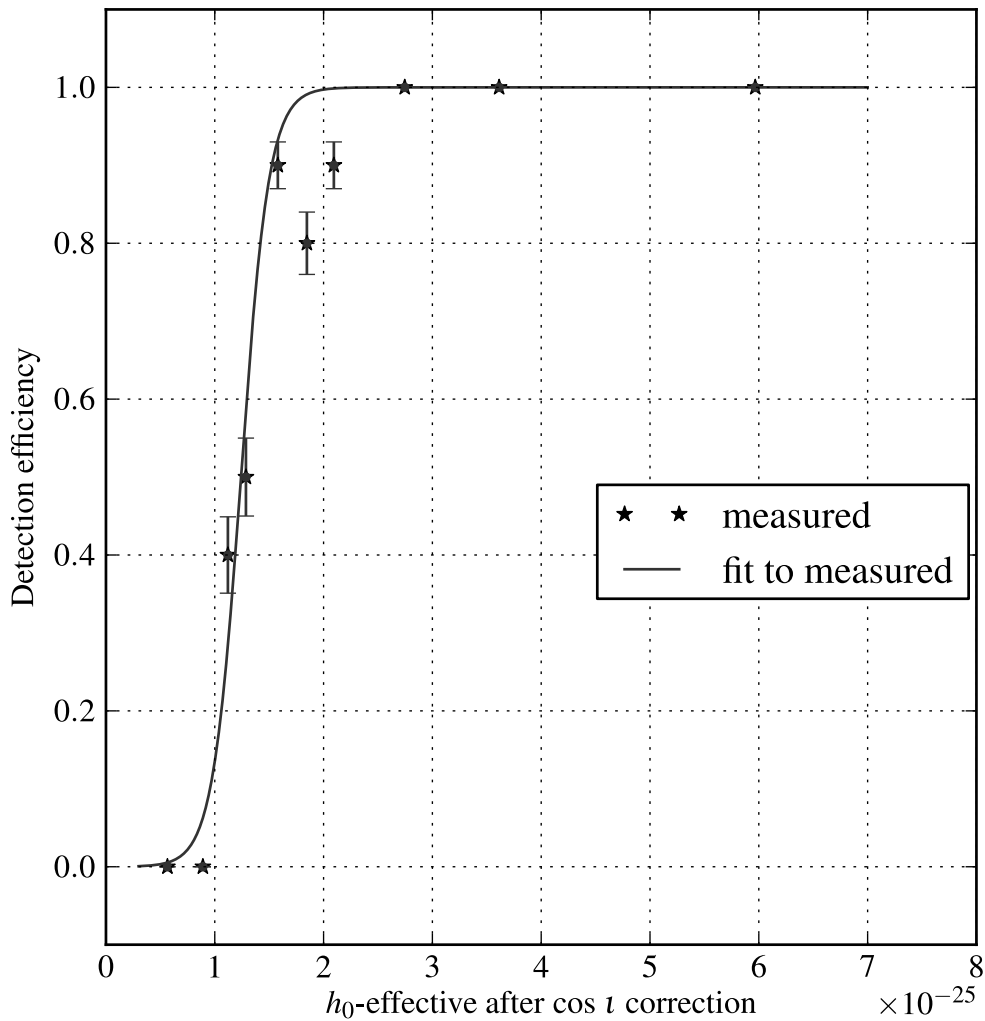}
\caption{Detection efficiency for open and closed signals.
Because only 100 signals are in the data set, this curve has fluctuations and large $1$-$\sigma$ error bars.
Moreover, the binomial error for these error bars is zero in bins where no or all signals are detected, which is not necessarily realistic.
The 95\% level is just below $2 \times 10^{-25}$ (before rescaling factors of 1.74 and 1.11), corroborating better techniques of estimating the upper limit.
\label{fig:detectionvsheffective}}
\end{center}
\end{figure}

To measure detection efficiency, we calculate the $h_\mathrm{eff}$ for all signals and find the average detection rate for a given $h_\mathrm{eff}$. 
Binomial uncertainty is also calculated and each 1-$\sigma$ deviation (per 5-signal bin) is graphed in Figure~\ref{fig:detectionvsheffective}, which shows a least-squares sigmoid fit.
This detection efficiency curve maps from strain to probability of detection.
Next we would like an upper limit function that takes a probability as an input and returns a strain that, with the given probability, is no less than the actual strain.

\begin{figure}
\begin{center}
\includegraphics[trim=0 0 0 0, clip, keepaspectratio,height=0.48\paperheight]{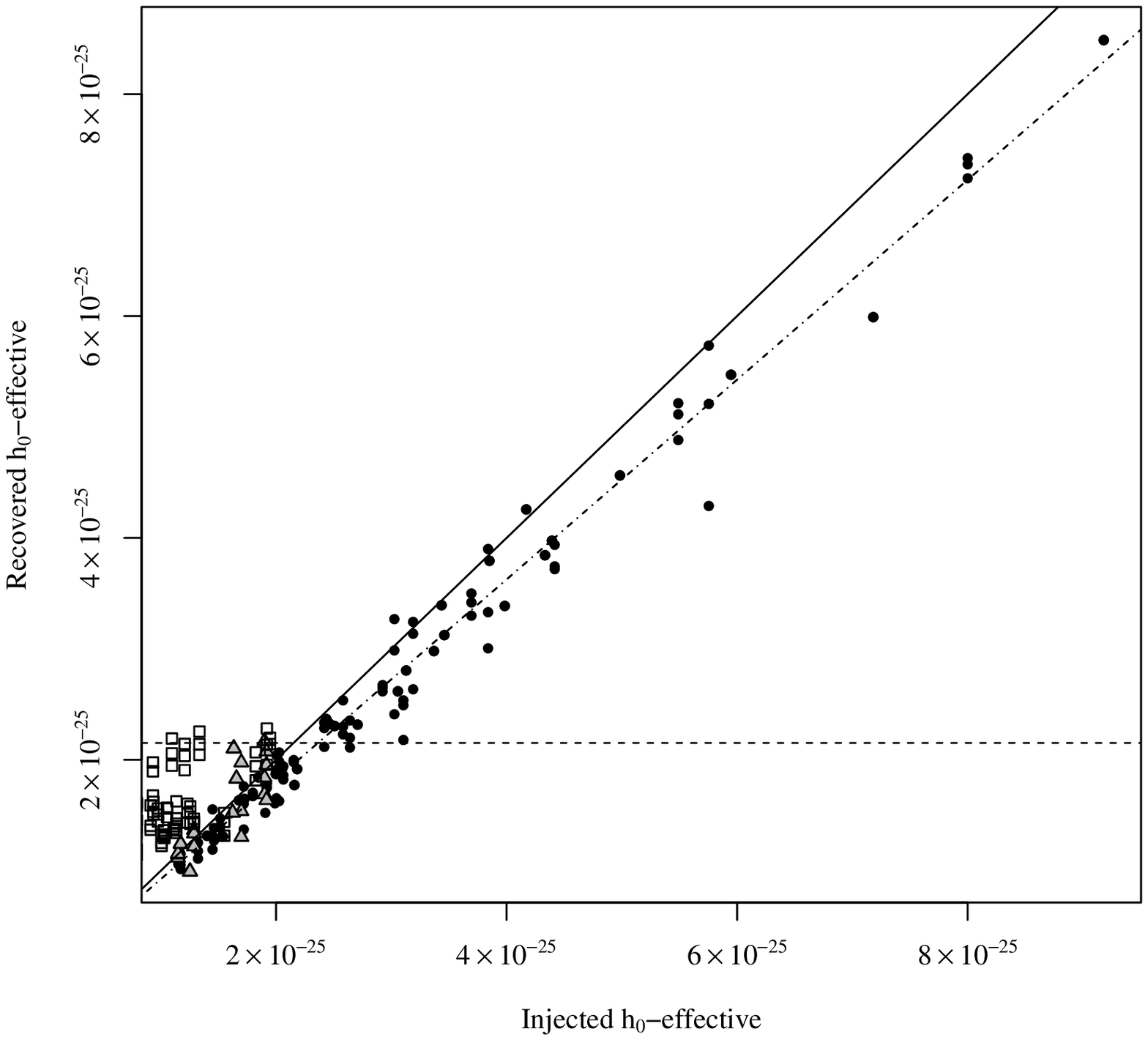}
\caption{Detections and upper limit determination, with all 100 simulated signals.
Injections are seen in three (black circle), one (gray triangle), or zero (open box) observatory pairs and plotted in recovered strain versus effective circular strain injected.
The most significant template is shown, in each band, in each observatory, whether detected or not.
(There are no injections seen with two detection pairs, because this plot shows only the loudest outlier from each 5 Hz band; if some injection were seen in two and not three pairs, it would mean two distinct coincidences were seen, only one of which would be the loudest).
We identify a shelf of non-detected signals that are 95\% covered by an upper limit about $2.19 \times 10^{-25}$. This number, when corrected, yielded the upper limit of 1.74*1.11*$2.19 \times 10^{-25}$ = $4.23 \times 10^{-25}$. 
The unity-slope black line is shown to ascertain whether a further empirical rescaling factor is needed to match the dashed-and-dotted least-squares linear fit (it is: constant 1.11).
The zero-slope horizontal dashed line is shown to indicate the ninety-five percent confidence upper limit in the absence of detection.
\label{fig:hrecoveredvsheffectivefullul}}
\end{center}
\end{figure}

To characterize the upper limit, we plot the distribution of $h_\mathrm{rec}$ versus injected $h_\mathrm{eff}$ 
in Figure~\ref{fig:hrecoveredvsheffectivefullul}. 
We verify that 95\% of non-detected open signals are covered by a naive upper limit of $h_\mathrm{rec} = 2.19 \times 10^{-25}$. 
This claim does not rely on binning but rather on the sampling density of the injections.
This number, when corrected by the $\rho_R$ rescaling factor of $1.11$ and $\cos \iota$ correction $\rho_\mathrm{\cos\iota}$ of 1.74, yields the upper limit of $(1.74)\times(1.11)\times2.19 \times 10^{-25}$ $=$ $4.23 \times 10^{-25}$.
Because of the flat noise floor, this is reported as a single upper limit for any non-detections in the MDC.

Frequency-dependent upper limits are well under development for actual observations.
Sigmoid fits to the detection efficiency of a set of injections into real data are generated, one fit per frequency band.
Upper limits at a given confidence can then be taken as the $h_0$ that yields a detection efficiency equal to that confidence.
This advancement post-dates the MDC and is planned for future applications.

\section{Conclusions}

\subsection{MDC results}

TwoSpect analyses applied to the MDC data set detect more stars than the Radiometer, Sideband, or Polynomial pipelines; only the CrossCorr algorithm found more signals~\cite{ScoX1MDC2015PRD}.
Each detection includes an estimate of $a \sin i$, which is not produced by Radiometer, Sideband, or Polynomial.
The MDC did not model the spin-wandering of the neutron star that is expected in real data, although participants were told to assume its presence, and spin-wandering is planned for future MDCs.
TwoSpect is also theoretically highly robust against spin-wandering.
This method has already been applied to real data~\cite{GoetzTwoSpectResults2014}, though not using the directed search in a fully-templated mode.
This experience validates the program's robustness with respect to non-Gaussian data artifacts.
In all,
34 of 50 closed (and 31 of 50 open) signals are detected, and
$f$, $a \sin i$, and $h_0$ are estimated.
Strain upper limits of $4.23\times 10^{-25}$ noise of in $4 \times 10^{-24}$ strain Hz$^{1/2}$ are determined for the 16 non-detected, closed signals.
Although the distribution of $h_0$ values in the MDC was astrophysically optimistic, the MDC validated our ability to claim detections and recover orbital and GW parameters accurately.

\subsection{Future directed CW binary searches}
Algorithms such as TwoSpect are designed to find astrophysically-plausible strain from LMXBs.
Torque-balance arguments suggest that strain could exceed $10^{-25}$ for Sco X-1 if it rotates at low frequencies.

The previously-published all-sky search in a year of S6 data set an overall upper limit for circular polarization of $2.3\times 10^{-24}$ at 217 Hz~\cite{GoetzTwoSpectResults2014}, set for an H1 amplitude spectral density of $2.0\times10^{-23}$ Hz$^{-1/2}$.
For random polarization, a multiplicative factor of $2.2$ was applied, for an upper limit of $5.1 \times 10^{-24}$.
This corresponds to a sensitivity depth (factor below the noise floor: ~\ref{mathematical-appendix-methods}) of 8.7 Hz$^{-1/2}$ for circular polarization but 4.0 Hz$^{-1/2}$ for random.
Extrapolating to Advanced LIGO design sensitivities of $4\times10^{-24}$ Hz$^{-1/2}$, this implies an upper limit around $5\times10^{-25}$ for circular polarization and $1\times 10^{-24}$ for random.
However, the all-sky paper included an opportunistic search for Sco X-1 on a narrow frequency range (20 to 57.25 Hz), setting a random polarization limit of $2\times 10^{-23}$ at 57 Hz in an L1 amplitude spectral density $1.8 \times 10^{-22}$ Hz$^{-1/2}$, a depth of 9 Hz$^{-1/2}$.
This opportunistic search used 1800-s SFTs; longer SFT durations increase theoretical sensitivity.
Significant improvement comes from focusing on one sky location; this can be viewed as a reduced trials factor.
The directed search demonstrated in this paper achieved an $4.23\times 10^{-25}$ upper limit in simulated data at design sensitivity, a depth of 9.5 Hz$^{-1/2}$.
It achieved this depth despite being tested with shorter (360-s and 840-s) SFTs. 
The directed search is also scalable over a much wider parameter range, like the all-sky method over which it gains twofold in sensitivity.

While real data complications may worsen this limit, several simplified and conservative steps were taken.
The limit may improve with the enhancements now under developement, when fully tested with injections as in the all-sky search.
Even now, the method in this paper is more sensitive for random polarization than the all-sky method is for optimal, circular polarization.
Additional improvements to the algorithm, such as coherent SFT summing, have been developed~\cite{TwoSpectCoherentGoetz2015arxiv} and could further improve this limit in the future, pushing toward the torque-balance strain.

Directed TwoSpect analyses have been demonstrated in this paper.
Comprehensively covering the parameter space of Sco X-1 at full sensitivity with the directed search, instead of hierarchically as before, does increase the probability of detection and improve upper limits.
When detections do occur, the ability to determine the frequency and projected semi-major axis of the neutron star in the binary system will prove highly informative.
Analyses of real data for signals from Sco X-1 and additional neutron stars in binary systems, such as XTE J1751-305, are underway.
In the long term, we hope that the discovery of gravitational waves from neutron stars in LMXBs will provide a firm link between our observations and electromagnetic astronomy.

\section*{Acknowledgments}

This work was partly funded by National Science Foundation grants NSF PHY 1205173 and NSF PHY 1505932.
These investigations have taken place as part of the LIGO Scientific Collaboration.
The Mock Data Challenge for Sco X-1 was organized by Chris Messenger.
Thanks to Maria Alessandra Papa for providing extensive guidance and support, as well as to John Whelan for thorough proofreading, along with our referees for their helpful comments.
Code for this paper is available online~\cite{LALAPPSrepo}.
This document bears LIGO DCC number P1500037.

\appendix
\section{Mathematical details}
\label{mathematical-appendix-methods}
\setcounter{section}{1}

\subsection{Sensitivity depth}

The relative sensitivity of a search can be quantified in terms of
sensitivity depth $D(f)$~\cite{BehnkeGalacticCenter2015,LeaciPrixDirectedFStatPRD},

\begin{equation}
D(f) \equiv \frac{S_H^{1/2}(f)}{h_0(f)},
\end{equation}

\noindent for noise power spectral density $S_H$ at frequency $f$ and the GW strain $h_0(f)$ recoverable there.
The depth generally depends on the observation time, because integrated signal-to-noise grows.
This concept allows us to compare methods across data sets and extrapolate future performance.

\subsection{Number of templates}

Equation~\ref{N_template_simple} is the integrated template density over parameter space dimensions.
While we are interested astrophysically in $a \sin i$, the observable $\Delta f_\mathrm{obs}$ governs template placement.
The equation decomposes into two number densities, $Q_\mathrm{frequency}$ and $Q_\mathrm{mod}$, and corresponding dimension length intervals, $L_\mathrm{frequency}$, $L_\mathrm{mod}$:

\begin{equation}
N_{\mathrm{template}} = {\displaystyle \Sigma}_{\mathrm{(frequency)}} {\displaystyle \Sigma}_{\mathrm{(mod)}} (L_\mathrm{frequency}\times Q_\mathrm{frequency}) (L_\mathrm{mod}\times Q_\mathrm{mod}),
\label{templates-explained}
\end{equation}

\noindent where $Q_\mathrm{frequency}$ is the inverse of the template spacing in frequency, which is $1/(2 T_\mathrm{coh})$, so $Q_\mathrm{frequency} = 2 T_\mathrm{coh}$.
Index the frequency dimension by $j$ bands.
One step $j$ is made per band, width $f_\mathrm{bw}$, so the length interval $L_\mathrm{frequency} = f_\mathrm{bw}$.
Thus

\begin{equation}
L_\mathrm{frequency}\times Q_\mathrm{frequency} = (f_\mathrm{bw}) \times (2 T_\mathrm{coh}).
\label{density-in-frequency}
\end{equation}

With fixed frequency bands, the number of templates per band does not change.
The width in modulation depth, however, depends on $a \sin i$ and the frequency $f_j = (f_\mathrm{min} + j f_\mathrm{bw})$.
Each template of $\Delta f_\mathrm{obs}$ is indexed by $k$. 
Analogous to before, $Q_{mod} = 1/(4 T_\mathrm{coh})$, so

\begin{eqnarray}
L_\mathrm{mod} \times Q_\mathrm{mod} &= \left(\Delta f_{\mathrm{mod, }~j,k}\right) \times \left(4 T_\mathrm{coh}\right),\\
         &= \left(\frac{2 \pi f_j}{cP} (a \sin i)_k\right) \times \left(4 T_\mathrm{coh}\right),
\label{density-in-mod}
\end{eqnarray}

\noindent when $a \sin i$ is given in light-seconds.
Substituting Equation~\ref{density-in-frequency} into Equation~\ref{templates-explained}, that term depends on neither the frequency nor modulation depth index, and so pulls out in front of the sums.
Equation~\ref{density-in-mod} depends on $k$; the sum $\Sigma_\mathrm{(mod)}$ is evaluated from $(a \sin i)_k = (a \sin i - 3 \sigma_{a \sin i})$ to $(a \sin i + 3 \sigma_{a \sin i})$ in practice, for an integrated length of $6 \sigma_{a \sin i}$.

Combining these elements, including indexing frequency band steps by $j$,

\begin{equation}
N_{\mathrm{template}} = \left(f_\mathrm{bw} \times 2 T_\mathrm{coh}\right) \Sigma_{j} \frac{2 \pi (f_\mathrm{min} + j f_\mathrm{bw})}{cP} 6 \sigma_{a \sin i} \times 4 T_\mathrm{coh}.
\end{equation}

\noindent Writing the limits of the sum of frequency bands, in addition to inserting ones to account for the edges of each sum, yields Equation~\ref{N_template_simple}.

\subsection{Test statistic calculation}

The construction of this $R$-statistic can be described in several steps.
The most important points from the original methods paper~\cite{GoetzTwoSpectMethods2011} are reiterated here, with some clarification.
Science/observing runs are first parcelled into overlapping short Fourier transforms (SFTs), performed in the detector frame.
The SFTs have typical coherence time $T_\mathrm{coh}$ (referred to as $T_\mathrm{SFT}$ in newer publications~\cite{TwoSpectCoherentGoetz2015arxiv}) ranging from 60 s to 1800 s, depending on the hypothesized time-derivative of neutron star frequency~\cite{GoetzTwoSpectMethods2011}.
The total number of SFTs with 50\%-overlap for an observing time $T_\mathrm{obs}$ is $N$,

\begin{equation}
N = \mathrm{floor}\left(\frac{2 T_\mathrm{obs}}{T_\mathrm{coh}}\right) - 1.
\end{equation}

\noindent SFT number in the observing run are indexed by $n \in [0,\ldots,N-1]$; SFT frequency bin is indexed by $k \in [0,\ldots,K -1]$, where $K = T_\mathrm{coh} f_N$ for a Nyquist frequency $f_N$ and only positive frequencies $k = T_\mathrm{coh} f$ are used.
Thus the transformation from time series to SFTs is a map from $x(t)$ ($=h(t) + n(t)$, signal strain plus noise) to $\tilde{x}^n_k$.

This array of $n$ still depends on detector time $t$, and the analysis is to be done in Solar System Barycenter (SSB) time $t_\mathrm{SSB}$.
Travel from the source to SSB introduces an overall phase shift; uncertainty in the distance and proper motion is systematic and the same for gravitational and electromagnetic observations.
Detector time is recorded in GPS time, running parallel with Terrestrial Time (TT), and SSB time runs parallel with Barycentric Dynamical Time (TDB).
SSB time corrects $t$ by $\delta t_R$ for relativistic effects.
Another overall phase shift is caused by  Roemer delay $\Delta t_R$, the dot product of $\hat{n}/c$ from the SSB to the sky location of interest with $\vec{r}$ from the SSB to the detector~\cite{Jaranowski1998,GoetzTwoSpectMethods2011,TwoSpectCoherentGoetz2015arxiv}.
Barycentering detector-frame data is equivalent to resampling in $\tau(t) = t_\mathrm{SSB}(t) + \Delta t_R(t)$,

\begin{equation}
\tau(t) 
 = t + \frac{\vec{r}(t) \cdot \hat{n}}{c} + \delta t_R.
\label{barycentering_time_domain}
\end{equation}

Each SFT frequency bin $\tilde{x}^n_k$ is Doppler shifted to a frequency bin in the SSB frame corresponding to the sky location, frequency, and time $t$ of the midpoint of the SFT $n$ under investigation: $k(f(\alpha,\delta,t)) \rightarrow k(f_\mathrm{SSB}(\tau))$.
This barycentering procedure corresponds to the time-domain Equation~\ref{barycentering_time_domain}.
Henceforth, barycentering is implicit in the $k$ index.

Define the power $P^n_k = 2|\tilde{x}^n_k|^2/T_\mathrm{SFT}$.
Let $\left<P_k\right>^n$ be the expected (estimated from a running mean over nearby $n$) noise-only power in a frequency bin $k$ for SFT $n$.
Also let $F^2_n \equiv F_{+,n}^2 + F_{\times,n}^2$ for the antenna pattern at the chosen sky location and SFT $n$ -- taking this equal-weighted sum of $F_+$ and $F_{\times}$ polarization components implies an assumption of circular polarization.
Then the estimated power in a given bin $\tilde{P}^n_k$ is normalized such that random, white, Gaussian noise will have an expectation value of 1~\cite{GoetzTwoSpectMethods2011}:

\begin{equation}
\tilde{P}^n_k = \frac{F_n^2 (P_k^n - \left<P_k\right>^n)}{(\left<P_k\right>^n)^2}\left[\sum\limits_{n'}^N \frac{F_{n'}^4}{(\left<P_k\right>^{n'})^2} \right]^{-1}.
\label{equation_with_antenna_pattern}
\end{equation}

Then each row of barycentered frequency bins $k$ is treated as a time series in $n$.
Power for bin $k$ in that time series is Fourier transformed by $\mathcal{F}_{f'}$ into $Z_k(f')$, where $f'$ is the second Fourier transform frequency.
During the transform, the background noise power $\lambda(f')$ is estimated from the noise in the SFTs, assuming the noise is Gaussian.
This second Fourier power $Z_k(f')$ follows a $\chi^2$ distribution with 2 degrees of freedom and mean 1.0, is proportional to $h^4$, and is constructed by

\begin{equation}
Z_k(f') = \frac{\left| \mathcal{F}_{f'} [\tilde{P}^n_k]  \right|^2}{\left< \lambda(f') \right>}.
\label{second_Fourier_power}
\end{equation}

\noindent When re-indexed by sorted template weight, $i$, $Z_k(f')$ becomes $Z_i$ in Equation~\ref{TwoSpect_R_statistic}.
Extensive discourse on the details of these calculations, as well as the estimation of background and calculation of template weights, is found is the original TwoSpect methods paper~\cite{GoetzTwoSpectMethods2011}, and a paper on coherent addition of SFTs rigorously derives the SFT power by including Dirichlet kernel terms.
Given the second Fourier power, we calculate the $R$-statistic its $p$-value, from which we seek to make a detection.

\bibliographystyle{iopart-num}
\bibliography{bibliography}

\end{document}